\documentclass[11pt]{article}
\usepackage{amsmath}
\usepackage{amsfonts}
\usepackage{amssymb}
\usepackage{graphicx}
\usepackage{dcolumn}
\usepackage{bm}

\input{tcilatex}

\begin{document}

\title{On the problem of \textit{Zitterbewegung}\\
of the Dirac electron
\thanks{Presented at Research Workshop of the Israel Science Foundation 
\textit{Correlated Electrons at High Magnetic Fields}, Ein-Gedi/Holon, 
Israel, 19-23 December 2004} }
\author{\textnormal{Daniel Sepunaru} \\
\\
{\small {\textit{Research Center for Quantum Communication Engineering,}} }\\
{\small {\textit{Holon Academic Institute of Technology, 52 Golomb Str.,
Holon 58102, Israel}} }\\
{\small {\textit{e-mail: danielsepunaru@walla.co.il}}}\\
{\small {\textnormal{Received 6 July 2004, revised 23 September 2004, accepted 8 October 2004}}}}
\date{}
\maketitle

\begin{abstract}
Reformulation of Dirac equation in terms of real quadratic division algebra
of quaternions is given. Similar equations with different mass term are
identified as suitable for description of free propagating quark motion. The
complete orthonormal set of the positive-energy plane wave solutions is
presented. Therefore, \textit{Zitterbewegung} phenomenon is absent in this
formulation. The probability current is proportional to the momentum, as in
standard Schr\"{o}dinger wave mechanics.\newline
\end{abstract}

\section{Introduction}

In this paper I will discuss an old, well-known and apparently
\textquotedblleft insignificant\textquotedblright\ problem that lies in the
foundation of the relativistic quantum mechanics or, more precisely, in the
foundation of the quantum electrodynamics. A story starts with the
celebrated paper of P.A.M. Dirac \cite{1} that describes relativistic motion
of a free propagating electron, and had tremendous success of
natural explanation of spin, correct non-relativistic limit, correct
coupling with the external magnetic field, correct gyromagnetic ratio and,
finally, prediction of positron.

The mathematical formalism used requires presence of the negative kinetic
energy states in order to obtain the complete orthonormal set of the
linearly independent fundamental solutions of the suggested equation.
Indeed, such states do not make sense from the physical point of view.

Since the Dirac equation is written in the Hamiltonian form, it allows us to
work in the Heisenberg representation and determine directly whether or not
a given observable is a constant of motion. Then we find, for example, that
the momentum is a conserved quantity as it should be. However, the orbital
momentum, as well as the spin, are not conserved separately and only the sum
of them is a constant of motion. Conventionally, the spin is associated with
the internal degree of freedom of the electron and therefore apparently has
nothing to do with isotropy of the space-time continuum (indeed, if we
assign to the quantum mechanical space-time point internal algebraic
structure, then this will be rather naturally expected result). Even more
surprising result \cite{2} is obtained if we consider velocity $\dot{\vec{x}}
$ in the Dirac formulation. Instantaneous group velocity of the electron has
only values $\pm c$ in spite of the non-zero rest mass of electron. In
addition, velocity of a free moving electron is not a constant of motion.

An analytic solution for the coordinate operator of a free propagating
electron was found by E. Schr\"{o}dinger \cite{3}. It turns out that in
addition to the uniform rectilinear motion consistent with the classical
electrodynamics, the Dirac electron executes oscillatory motion, which E.
Schr\"{o}dinger called \textit{Zitterbewegung}. Let us recall that entire
non-relativistic quantum mechanics was raised in order to explain the
absence of radiation during the oscillatory motion of the electron bounded
by the electric potential of the nucleus. Therefore, the Dirac theory of
electron contains a definite prediction that the free moving electron will
loose all a kinetic energy through electromagnetic radiation \cite{4}.

It is rather surprising that the \textit{Zitterbewegung Problem} attracted
only sporadic \cite{5, 6} attention during years of development of the
theory of quantum fields and efforts to achieve the unification of all
fundamental interactions. It was demonstrated \cite{6, 7} that the \textit{%
Zitterbewegung} oscillations are due solely to interference between the
positive- and negative- energy components in the wave packet. The \textit{%
Zitterbewegung} is completely absent for a wave packet made up exclusively
of positive energy plane wave solutions. It is clear from the above analysis
that if one achieves the reformulation of the Dirac equation such that the
complete orthonormal set of linearly independent solutions will contain only
positive energy states then the \textit{Zitterbewegung} oscillations will
disappear. Indeed, the charge-conjugated solutions, associated with the
positron, must be retained.

It has been known for a long time that the algebraic structure of Dirac
equation is closely related to the quadratic normal division algebra of
quaternions. Here we suggest a quaternionic reformulation of the Dirac
equations \cite{8}, as well as an additional set of similar equations
suitable for description of the free propagating quark motion. The main
effort is made to obtain equations with the intrinsic $SU\left( 2\right)
\,\,\otimes U\left( 1\right) $ local gauge invariance. In contrast with the
approach of S.L. Adler and others \cite{9, 10}, we consider the possibility
that the previously obtained quaternionic extension \cite{11} of the Hilbert
space description of quantum fields represents a consistent mathematical
framework for the electroweak unification scheme (a brief summary of
relevant results is given in the Appendix). It is obvious that in order to
achieve unification of all fundamental interactions, the algebraic extension
beyond the quaternions is needed. We demonstrate that mathematical structure
of the obtained equations of motion suggests that the required extension may
proceed through wave functions which possess three and seven phases, whereas
the scalar product remains complex. In that case the examples of
nonextendability to octonionic quantum mechanics \cite{9} are not valid.

\section{Equations of motion for fundamental fermions}

Let us consider the algebraic structure of the Dirac equation. The problem
is to achieve factorization of the energy-momentum relation

\begin{equation}  \label{eq1}
E^2\,\, = \,\,p^2c^2\,\, + \,\,m^2c^4
\end{equation}

\noindent in such a way, that the correspondent Hamiltonian is the generator
of Abelian translations in time, which is expressed by the Schr\"{o}dinger
equation

\begin{equation}  \label{eq2}
i\hbar \,\,\frac{\partial \psi }{\partial t}\,\, = \,\,H\psi .
\end{equation}

It was demonstrated by P.A.M. Dirac \cite{1} that in terms of the
two-dimensional commutative quadratic division algebra of complex numbers no
solution can be found. The problem requires intrinsically an extension of
the algebraic basis of the theory. The Dirac's solution of the problem,

\begin{equation}
i\hbar \,\,\frac{\partial \psi }{\partial t}\,\,=\,\,\frac{\hbar c}{i}{%
\alpha _{j}\frac{\partial \psi }{\partial x_{j}}}\,\,-\,\,\beta
\,\,mc^{2}\,\,\psi \,\,\equiv \,\,H\psi ,\text{ \ }\ \ \text{\ }j=1,2,3\text{%
\ }  \label{eq3}
\end{equation}

\newpage

\noindent uses the generators of the $C_{4}$ Clifford algebra:

\begin{equation}
\begin{array}{l}
{\alpha _{i}\alpha _{k}\,\,+\,\,\alpha _{k}\alpha _{i}\,\,=\,\,2\delta _{ik}}
\\ 
{\alpha _{i}\beta \,\,+\,\,\beta \alpha _{i}\,\,=\,\,0} \\ 
{\alpha _{i}^{2}\,\,=\,\,\beta ^{2}\,\,=\,\,1.}%
\end{array}%
\quad {i,k\,\,=\,\,1,2,3}  \label{eq4}
\end{equation}

However, such a drastic growth in algebra is only apparent. The true
physical content of the obtained result is expressed more distinctly if (\ref%
{eq3}) is written in the following form:

\begin{equation}
\begin{array}{l}
i\dfrac{1}{c}\,\,\dfrac{\partial \psi }{\partial t}\,\,+\,i\sigma _{j}\dfrac{%
\partial \psi }{\partial x_{j}}\,=\,\,\dfrac{mc}{\hbar }\phi \\ 
\\ 
-i\dfrac{1}{c}\,\,\dfrac{\partial \phi }{\partial t}\,\,+i\sigma _{j}\dfrac{%
\partial \phi }{\partial x_{j}}\,\,\,=\,\,-\dfrac{mc}{\hbar }\psi%
\end{array}%
\quad j=1,2,3  \label{eq5}
\end{equation}%
(we choose to work in the Weyl representation \cite{12} 
\begin{equation}
\begin{array}{l}
{\alpha _{i}\,\,=\,\,\left( {{\begin{array}{*{20}c} {\sigma _i } \hfill & 0
\hfill \\ 0 \hfill & { - \sigma _i } \hfill \\ \end{array}}}\right) } \\ 
\\ 
{\beta \,\,=\,\,\left( {{\begin{array}{*{20}c} 0 \hfill & { - 1} \hfill \\ {
- 1} \hfill & 0 \hfill \\ \end{array}}}\right) } \\ 
\\ 
{\Psi \,\,=\,\,\left( {{\begin{array}{*{20}c} \psi \\ \phi \\ \end{array}}}%
\right) }%
\end{array}%
\quad i=1,2,3  \label{eq6}
\end{equation}

\noindent by the reason, which will be explained below).

In a precise analogy with the physical content of Maxwell's equations, we
again have to deal with two mutually connected waves, which propagate
together in a space with constant velocity.

Now, the full basis of the certain algebra ($C_2 $ Clifford algebra) is
symmetrically used in (\ref{eq5}). However, it is assumed that $i$ in (\ref%
{eq5}) commutes with $\sigma _i $, that is the algebra is defined over the
field of complex numbers. Therefore, the algebraic foundation of this
formulation is based on an eight-dimensional non-division algebra. In
addition, it is customary in the applications to continue working with
complex numbers as an abstract algebra, but for the $C_2 $ Clifford algebra,
one makes use of a representation (Pauli matrices) introducing into the
theory an asymmetry, which has neither mathematical nor physical
justification.

Now, I will demonstrate that the algebraic foundation of (\ref{eq5}) may be
reduced to a four-dimensional real quadratic division algebra of quaternions
and that the structure of Dirac equations is intrinsically connected with
the functional-analytical structures mentioned above.

First of all, let us substitute

\begin{equation}  \label{eq7}
e_j \,\, = \,\, - i\sigma _j \,\,\,\,\,\,\,\,\,\,\,\,\,\,\,j\,\, = \,\,1,2,3
\end{equation}

\noindent into (\ref{eq5}). Then%
\begin{equation}
\begin{array}{l}
i\dfrac{1}{c}\,\dfrac{\partial \psi }{\partial t}\,\,-e_{j}\dfrac{\partial
\psi }{\partial x_{j}}\,\,\,=\,\,\dfrac{mc}{\hbar }\phi \\ 
\\ 
-i\dfrac{1}{c}\,\dfrac{\partial \phi }{\partial t}\,\,-\,e_{j}\dfrac{%
\partial \phi }{\partial x_{j}}\,\,\,=\,\,-\dfrac{mc}{\hbar }\psi%
\end{array}%
\quad j=1,2,3  \label{eq8}
\end{equation}

\noindent or, equivalently,

\begin{equation}
\begin{array}{l}
\left\{ {\left( {{\begin{array}{*{20}c} 0 \hfill & {e_0 } \hfill \\ {e_0 }
\hfill & 0 \hfill \\ \end{array}}}\right) \,\dfrac{1}{c}\,\dfrac{\partial }{%
\partial t}\,-\left( {{\begin{array}{*{20}c} {e_j } \hfill & 0 \hfill \\ 0
\hfill & { - e_j } \hfill \\ \end{array}}}\right) \dfrac{\partial }{\partial
x_{j}}}\right\} \,\left( {{\begin{array}{*{20}c} \psi \hfill \\ {\psi i}
\hfill \\ \end{array}}}\right) \\ 
\\ 
\text{ \ \ \ \ \ \ \ \ \ \ \ \ \ \ \ \ \ \ \ \ \ \ \ \ \ \ \ \ \ }=\dfrac{mc%
}{\hbar }\,\,\left( {{\begin{array}{*{20}c} {e_0 } \hfill & 0 \hfill \\ 0
\hfill & { - e_0 } \hfill \\ \end{array}}}\right) \left( {{%
\begin{array}{*{20}c} \phi \hfill \\ {\phi i} \hfill \\ \end{array}}}\right)
\\ 
\\ 
\left\{ {-\left( {{\begin{array}{*{20}c} 0 \hfill & {e_0 } \hfill \\ {e_0 }
\hfill & 0 \hfill \\ \end{array}}}\right) \dfrac{1}{c}\dfrac{\partial }{%
\partial t}-\left( {{\begin{array}{*{20}c} {e_j } \hfill & 0 \hfill \\ 0
\hfill & { - e_j } \hfill \\ \end{array}}}\right) \dfrac{\partial }{\partial
x_{j}}}\right\} \,\left( {{\begin{array}{*{20}c} \phi \hfill \\ {\phi i}
\hfill \\ \end{array}}}\right) \, \\ 
\\ 
\text{ \ \ \ \ \ \ \ \ \ \ \ \ \ \ \ \ \ \ \ \ \ \ \ \ \ \ \ \ }=-\dfrac{mc}{%
\hbar }\,\left( {{\begin{array}{*{20}c} {e_0 } \hfill & 0 \hfill \\ 0 \hfill
& { - e_0 } \hfill \\ \end{array}}}\right) \left( {{\begin{array}{*{20}c}
\psi \hfill \\ {\psi i} \hfill \\ \end{array}}}\right)%
\end{array}%
\quad j=1,2,3  \label{eq9}
\end{equation}

\noindent and

\begin{equation}
\begin{array}{l}
\left\{ {\left( {{\begin{array}{*{20}c} 0 \hfill & { - e_0 } \hfill \\ {e_0
} \hfill & 0 \hfill \\ \end{array}}}\right) \,\dfrac{1}{c}\,\dfrac{\partial 
}{\partial t}\,-\left( {{\begin{array}{*{20}c} {e_j } \hfill & 0 \hfill \\ 0
\hfill & {e_j } \hfill \\ \end{array}}}\right) \dfrac{\partial }{\partial
x_{j}}}\right\} \,\left( {{\begin{array}{*{20}c} \psi \hfill \\ { - \psi i}
\hfill \\ \end{array}}}\right) \\ 
\\ 
\text{ \ \ \ \ \ \ \ \ \ \ \ \ \ \ \ \ \ \ \ \ \ \ \ \ \ \ \ \ \ \ }=\dfrac{%
mc}{\hbar }\,\,\left( {{\begin{array}{*{20}c} {e_0 } \hfill & 0 \hfill \\ 0
\hfill & {e_0 } \hfill \\ \end{array}}}\right) \left( {{%
\begin{array}{*{20}c} \phi \hfill \\ { - \phi i} \hfill \\ \end{array}}}%
\right) \\ 
\\ 
\left\{ {-\left( {{\begin{array}{*{20}c} 0 \hfill & { - e_0 } \hfill \\ {e_0
} \hfill & 0 \hfill \\ \end{array}}}\right) \dfrac{1}{c}\,\dfrac{\partial }{%
\partial t}-\left( {{\begin{array}{*{20}c} {e_j } \hfill & 0 \hfill \\ 0
\hfill & {e_j } \hfill \\ \end{array}}}\right) \dfrac{\partial }{\partial
x_{j}}}\right\} \,\left( {{\begin{array}{*{20}c} \phi \hfill \\ { - \phi i}
\hfill \\ \end{array}}}\right) \\ 
\, \\ 
\text{ \ \ \ \ \ \ \ \ \ \ \ \ \ \ \ \ \ \ \ \ \ \ \ \ \ \ \ \ }=-\dfrac{mc}{%
\hbar }\,\left( {{\begin{array}{*{20}c} {e_0 } \hfill & 0 \hfill \\ 0 \hfill
& {e_0 } \hfill \\ \end{array}}}\right) \left( {{\begin{array}{*{20}c} \psi
\hfill \\ { - \psi i} \hfill \\ \end{array}}}\right) .%
\end{array}%
\quad j=1,2,3  \label{eq10}
\end{equation}

Notice that the states have the form (A1) and all operators have the form
(A11) and (A12).

Besides that, an additional (and only one) mass term is allowed:

\begin{equation*}
\begin{array}{l}
\left\{ {\left( {{\begin{array}{*{20}c} 0 \hfill & {e_0 } \hfill \\ {e_0 }
\hfill & 0 \hfill \\ \end{array}}}\right) \dfrac{1}{c}\,\dfrac{\partial }{%
\partial t}-\left( {{\begin{array}{*{20}c} {e_j } \hfill & 0 \hfill \\ 0
\hfill & { - e_j } \hfill \\ \end{array}}}\right) \dfrac{\partial }{\partial
x_{j}}}\right\} \,\left( {{\begin{array}{*{20}c} \psi \hfill \\ {\psi i}
\hfill \\ \end{array}}}\right) \\ 
\\ 
\text{ \ \ }=\left\{ {\dfrac{m_{1}c}{\hbar }\,\,\left( {{%
\begin{array}{*{20}c} {e_0 } \hfill & 0 \hfill \\ 0 \hfill & { - e_0 }
\hfill \\ \end{array}}}\right) \,\,+\,\,\dfrac{m_{2}c}{\hbar }\,\,\left( {{%
\begin{array}{*{20}c} 0 \hfill & {e_0 } \hfill \\ {e_0 } \hfill & 0 \hfill
\\ \end{array}}}\right) }\right\} \,\left( {{\begin{array}{*{20}c} \phi
\hfill \\ {\phi i} \hfill \\ \end{array}}}\right) \\ 
\\ 
\left\{ {-\left( {{\begin{array}{*{20}c} 0 \hfill & {e_0 } \hfill \\ {e_0 }
\hfill & 0 \hfill \\ \end{array}}}\right) \,\dfrac{1}{c}\,\dfrac{\partial }{%
\partial t}\,\,-\,\left( {{\begin{array}{*{20}c} {e_j } \hfill & 0 \hfill \\
0 \hfill & { - e_j } \hfill \\ \end{array}}}\right) \dfrac{\partial }{%
\partial x_{j}}\,}\right\} \,\,\left( {{\begin{array}{*{20}c} \phi \hfill \\
{\phi i} \hfill \\ \end{array}}}\right) \, \\ 
\\ 
\text{ \ \ }=\left\{ {-\dfrac{m_{1}c}{\hbar }\,\,\left( {{%
\begin{array}{*{20}c} {e_0 } \hfill & 0 \hfill \\ 0 \hfill & { - e_0 }
\hfill \\ \end{array}}}\right) \,\,+\,\,\dfrac{m_{2}c}{\hbar }\,\,\left( {{%
\begin{array}{*{20}c} 0 \hfill & {e_0 } \hfill \\ {e_0 } \hfill & 0 \hfill
\\ \end{array}}}\right) }\right\} \,\,\left( {{\begin{array}{*{20}c} \psi
\hfill \\ {\psi i} \hfill \\ \end{array}}}\right) \\ 
\end{array}%
\text{ \ \ \ \ }j=1,2,3
\end{equation*}

\begin{equation}
\begin{array}{l}
\left\{ {\left( {{\begin{array}{*{20}c} 0 \hfill & { - e_0 } \hfill \\ {e_0
} \hfill & 0 \hfill \\ \end{array}}}\right) \dfrac{1}{c}\,\dfrac{\partial }{%
\partial t}-\left( {{\begin{array}{*{20}c} {e_j } \hfill & 0 \hfill \\ 0
\hfill & {e_j } \hfill \\ \end{array}}}\right) \dfrac{\partial }{\partial
x_{j}}}\right\} \,\left( {{\begin{array}{*{20}c} \psi \hfill \\ { - \psi i}
\hfill \\ \end{array}}}\right) \, \\ 
\\ 
\text{ \ \ }=\left\{ {\dfrac{m_{1}c}{\hbar }\,\,\left( {{%
\begin{array}{*{20}c} {e_0 } \hfill & 0 \hfill \\ 0 \hfill & {e_0 } \hfill
\\ \end{array}}}\right) \,\,+\,\,\dfrac{m_{2}c}{\hbar }\,\,\left( {{%
\begin{array}{*{20}c} 0 \hfill & { - e_0 } \hfill \\ {e_0 } \hfill & 0
\hfill \\ \end{array}}}\right) }\right\} \,\left( {{\begin{array}{*{20}c}
\phi \hfill \\ { - \phi i} \hfill \\ \end{array}}}\right) \, \\ 
\\ 
\left\{ {-\left( {{\begin{array}{*{20}c} 0 \hfill & { - e_0 } \hfill \\ {e_0
} \hfill & 0 \hfill \\ \end{array}}}\right) \,\,\dfrac{1}{c}\,\dfrac{%
\partial }{\partial t}\,\,-\,\left( {{\begin{array}{*{20}c} {e_j } \hfill &
0 \hfill \\ 0 \hfill & {e_j } \hfill \\ \end{array}}}\right) \dfrac{\partial 
}{\partial x_{j}}\,}\right\} \,\,\left( {{\begin{array}{*{20}c} \phi \hfill
\\ { - \phi i} \hfill \\ \end{array}}}\right) \, \\ 
\\ 
\text{ \ \ }=\left\{ {\dfrac{m_{1}c}{\hbar }\,\,\left( {{%
\begin{array}{*{20}c} {e_0 } \hfill & 0 \hfill \\ 0 \hfill & {e_0 } \hfill
\\ \end{array}}}\right) \,\,+\,\,\dfrac{m_{2}c}{\hbar }\,\,\left( {{%
\begin{array}{*{20}c} 0 \hfill & { - e_0 } \hfill \\ {e_0 } \hfill & 0
\hfill \\ \end{array}}}\right) }\right\} \,\,\left( {{\begin{array}{*{20}c}
\psi \hfill \\ { - \psi i} \hfill \\ \end{array}}}\right) .%
\end{array}%
\text{ \ \ \ }j=1,2,3  \label{eq11}
\end{equation}

It may be verified that the energy-momentum relation is not spoiled if one
defines

\begin{equation}
M\equiv \,\,\sqrt{m_{1}^{2}\,\,+\,\,m_{2}^{2}}.  \label{eq12}
\end{equation}

Here we are forced to consider masses as given phenomenological parameters.
If $m_{1}\,\,\neq \,\,0$, the presence of this additional term does not
increase the number of fundamental plane wave solutions of the equations (%
\ref{eq11}). Therefore, we will consider the equations (\ref{eq11}) with $%
m_{1}\,\,=\,\,0$ as a separate independent set and in order to maintain the
direct connection with the Dirac equations, will neglect the $m_{2}$ term in
the presence of the non-vanishing $m_{1}$ term.

Indeed, only two equations (\ref{eq11}) are independent:

\begin{equation}
\begin{array}{l}
\dfrac{1}{c}\,\,\dfrac{\partial \psi }{\partial t}i\,\,-\,e_{j}\,\dfrac{%
\partial \psi }{\partial x_{j}}\,\,=\,\,\dfrac{mc}{\hbar }\phi \\ 
\\ 
-\dfrac{1}{c}\,\,\dfrac{\partial \phi }{\partial t}i\,\,-e_{j}\,\dfrac{%
\partial \phi }{\partial x_{j}}\,\,=\,-\,\dfrac{mc}{\hbar }\psi .%
\end{array}%
\quad j=1,2,3  \label{eq13}
\end{equation}

The form (\ref{eq13}) is very convenient for the investigation of gauge
invariance group of the Dirac \noindent equations. The $U\,\,\left( 1\right) 
$ gauge invariance group from the right is generated by the transformations

\begin{equation}
\begin{array}{l}
\psi ^{\prime }\,\,=\,\,\psi \,z,\,\,\,\,\,\phi ^{\prime }\,\,=\,\,\phi \,z
\\ 
\\ 
z\,\,=\,\,a\,\,+\,\,bi,\,\,\,\,\,\left\vert z\right\vert \,\,=\,\,1,\text{ }%
a,\,\,b\,\,\text{are real numbers}.%
\end{array}
\label{eq14}
\end{equation}

Since for every pair of solutions $\left( {\psi ,\,\,\phi } \right)$ of the
linear differential equations, $\left( {\psi a,\,\,\phi a} \right)$ ($a$ is
a real number) is also a solution, it is always enough to show that the
particular transformation

\begin{equation}
{\begin{array}{*{20}c} {\psi '\,\, = \,\,\psi \,i} \\ \\ {\phi '\,\, =
\,\,\phi \,i} \\ \end{array}}\,\,\,{\begin{array}{*{20}c} \hfill \\ {\left(
{\rm{i.e.}\,\,\,\,\,a\,\, = \,\,0,\,\,\,\,\,\,\,\,\,b\,\, = \,\,1} \right)}
\hfill \\ \hfill \\ \end{array}}  \label{eq15}
\end{equation}

\noindent leaves the equations invariant.

Invariance of the equations (\ref{eq13}) with respect to this transformation
is obvious. Let us consider what is a left gauge invariance group of the
Dirac equations. Remember that (\ref{eq13}) are the equations for free
propagating waves and thus admit solutions of the form

\begin{equation}
\begin{array}{l}
\psi \,\,=\,\,U_{1}\,\,\exp \dfrac{-i\left( {Et\,\,-\,\,\vec{p}\vec{x}}%
\right) }{\hbar } \\ 
\\ 
\phi \,\,=\,\,U_{2}\,\,\exp \dfrac{-i\left( {Et\,\,-\,\,\vec{p}\vec{x}}%
\right) }{\hbar }.%
\end{array}
\label{eq16}
\end{equation}

Therefore, the $U\,\,\left( 1 \right)$ transformations

\begin{equation}
\begin{array}{l}
\psi ^{\prime }\,\,=\,\,z_{1}\psi ,\,\,\,\,\,\,\,\,\,\,\,\,\,\,\,\phi
^{\prime }\,\,=\,\,z_{1}\phi \\ 
\\ 
z_{1}\,\,=\,\,a\,\,+\,\,bi_{1},\,\,\,\,\,\,\,\,\,\left\vert {z_{1}}%
\right\vert \,\,=\,\,1 \\ 
\\ 
i_{1}\,\,\equiv \,\,\dfrac{e_{1}p_{1}\,\,+\,\,e_{2}p_{2}\,\,+\,%
\,e_{1}e_{2}p_{3}}{\left\vert \vec{p}\right\vert }%
\end{array}
\label{eq17}
\end{equation}

\noindent leave the equations (\ref{eq13}) invariant and constitute the left
gauge invariance group of the Dirac equations.

In order to see that, it is sufficient to show again that

\begin{equation}
\begin{array}{l}
\psi ^{\prime }\,\,=\,\,i_{1}\psi \\ 
\phi ^{\prime }\,\,=\,\,i_{1}\phi%
\end{array}
\label{eq18}
\end{equation}

\noindent is a solution of the equations (\ref{eq13}):

\begin{equation}
\begin{array}{l}
\dfrac{i_{1}U_{1}}{c}\left( {-iE}\right) \,\,i-\,e_{1}i_{1}U_{1}\,\,\left( {%
\,ip_{1}}\right) -\,e_{2}i_{1}U_{1}\,\left( {\,ip_{2}}\right)
-\,e_{1}e_{2}i_{1}U_{1}\,\left( {\,ip_{3}}\right) =\,mc\,\,i_{1}U_{2} \\ 
\\ 
\dfrac{i_{1}U_{2}}{c}\left( {iE}\right) \,\,i-\,e_{1}i_{1}U_{2}\,\,\left( {%
\,ip_{1}}\right) -\,e_{2}i_{1}U_{2}\,\left( {\,ip_{2}}\right)
-\,e_{1}e_{2}i_{1}U_{2}\,\left( {\,ip_{3}}\right) =-mc\,\,i_{1}U_{1}.%
\end{array}
\label{eq19}
\end{equation}

Then

\begin{equation}
\begin{array}{l}
\dfrac{i_{1}U_{1}E}{c}-\left( {e_{1}p_{1}+\,e_{2}p_{2}\,+\,e_{1}e_{2}p_{3}}%
\right) \,\,i_{1}U_{1}\,i=mc\,\,i_{1}U_{2} \\ 
\\ 
-\dfrac{i_{1}U_{2}E}{c}-\left( {e_{1}p_{1}+\,e_{2}p_{2}+\,e_{1}e_{2}p_{3}}%
\right) \,\,i_{1}U_{2}\,i=-mc\,\,i_{1}U_{1}.%
\end{array}
\label{eq20}
\end{equation}

By definition (see (\ref{eq17})),

\begin{equation}
e_{1}p_{1}\,+\,e_{2}p_{2}\,+\,e_{1}e_{2}p_{3}\,=\,i_{1}\,\,\left\vert \vec{p}%
\right\vert .  \label{eq21}
\end{equation}

Therefore,

\begin{equation}
\begin{array}{l}
i_{1}\,\,\left[ {\dfrac{U_{1}E}{c}\,-\,\left( {e_{1}p_{1}\,+\,e_{2}p_{2}\,+%
\,e_{1}e_{2}p_{3}}\right) \,U_{1}i}\right] \,=\,i_{1}\,\left( {mc\,U_{2}}%
\right) \\ 
\\ 
i_{1}\,\,\left[ {-\dfrac{U_{2}E}{c}-\,\left( {%
e_{1}p_{1}+e_{2}p_{2}+e_{1}e_{2}p_{3}}\right) \,U_{2}i}\right]
=\,\,i_{1}\,\left( {-mc\,U_{1}}\right) .%
\end{array}
\label{eq22}
\end{equation}

It is assumed in the Dirac equations \cite{ 13}, that

\begin{equation}  \label{eq23}
\left[ {i,e_j } \right]\,\, = \,\,0,\,\,\,\,\,\,\,\,\,j\,\, = \,\,1,2,3
\end{equation}

\noindent and hence the obtained gauge invariance group is $U\left( 1
\right)\,\, \otimes U\left( 1 \right)$. Now it becomes clear why the Dirac
equations allow us to incorporate an additional charge \cite{ 14} and turn
out to be suitable for the realization of the electroweak unification scheme 
\cite{ 15} without contradiction with the Aharonov-Bohm effect \cite{ 16}.

However, the group-theoretical content of this scheme \cite{ 15}, side by
side with the functional-analytical structures \cite{ 11}, suggests that the
left gauge invariance group should be larger $\left( {\mbox{at}\,\,%
\mbox{least}\,\,U\,\,\left( {1;q} \right)\,\, \cong \,\,SU\left( 2 \right)}
\right)$ and should not contain an Abelian invariant subgroup. The simplest
way to satisfy these requirements is to identify the Abelian groups (\ref%
{eq14}) and (\ref{eq17}) discussed above, that is to attach to the Dirac
equations the following form:

\begin{equation}
\begin{array}{l}
\dfrac{1}{c}\,\,\dfrac{\partial \psi }{\partial t}i\,\,-\,\,\dfrac{\partial
\psi }{\partial x_{j}}e_{j}\,\,=\,\,\dfrac{mc}{\hbar }\phi \\ 
\\ 
-\dfrac{1}{c}\,\,\dfrac{\partial \phi }{\partial t}i\,\,-\,\dfrac{\partial
\phi }{\partial x_{j}}e_{j}\,\,=\,-\,\dfrac{mc}{\hbar }\psi%
\end{array}%
\quad j=1,2,3  \label{eq24}
\end{equation}

\noindent and to drop the assumption (\ref{eq23}). Then the algebraic
foundation of the theory is reduced to a four-dimensional real quadratic
division algebra of quaternions.

The $U\left( 1 \right)$ right gauge invariance of the equations (\ref{eq24})
may be maintained~if

\begin{equation}  \label{eq25}
i\,\, = \,\,\frac{e_1 p_1 \,\, + \,\,e_2 p_2 \,\, + \,\,e_1 e_2 p_3 }{\left| 
\vec {p} \right|}
\end{equation}

\noindent and may be demonstrated exactly in the same way as (\ref{eq18}) - (%
\ref{eq22}).

Consequently, we have obtained the additional meaning for $i$, which appears
originally in the Schr\"{o}dinger equation. An algebra itself forms a vector
space, and the basis of algebra constitutes a suitable set of orthogonal
axes in that space, for example, a complex algebra may be considered as a
two-dimensional plane with the orthogonal directions 1 and $i$. In that
space $i$ standing in the left-hand side of the Schr\"{o}dinger equation
define the direction of the time translations, which form an Abelian group.
Therefore, the $U\left( 1\right) $ right gauge invariance of the equations (%
\ref{eq24}) leads us to the conclusion that these equations define (the
condition (\ref{eq25})) the direction of the time translations at the
three-dimensional quaternionic surface (the space of quantum mechanical
phases). Then, a possible physical interpretation is that, compared with a
classical relativistic particle, a quantum particle has not only its proper
time but, in addition, a proper direction of time. Perhaps, this may serve
as an explanation of why quantum equations of motion contain the first time
derivative, whereas the classical equations of motion are expressed in terms
of the second derivative.

We investigate now how our manipulations have affected the corresponding
solutions. As it is well known, the general solution of Dirac equation may
be formed as a linear combination of the four independent solutions, which
are four spinors with four components. Two of them are obtained for $E\,>0$
for two spin states $\,U_{2}^{(\ref{eq1})}\,=\left( {{%
\begin{array}{*{20}c}
 1 \hfill \\
 0 \hfill \\
\end{array}}}\right) $ and $\,U_{2}^{(\ref{eq2})}\,=\left( {{%
\begin{array}{*{20}c}
 0 \hfill \\
 1 \hfill \\
\end{array}}}\right) $, respectively. The other two we are forced to obtain
using $E\,\,$\TEXTsymbol{<} 0 since there are no other possibilities. They
correspond to the arbitrary choice of $\,U_{1}^{(\ref{eq3})}\,=\left( {{%
\begin{array}{*{20}c}
 1 \hfill \\
 0 \hfill \\
\end{array}}}\right) $ and $\,U_{1}^{(\ref{eq4})}\,=\left( {{%
\begin{array}{*{20}c}
 0 \hfill \\
 1 \hfill \\
\end{array}}}\right) $.

Let us check what happens in our quaternionic version of the Dirac equation.
In order to maintain connection with the original Dirac solutions, let us
form a complete orthonormal set of it:%
\begin{eqnarray}
\psi _{D}^{(\ref{eq1})}\,\, &=&\,\,\left( {{\begin{array}{*{20}c} {U_1
\,\,\left( \vec {p} \right)} \hfill \\ {U_1 \,\,\left( \vec {p} \right)i}
\hfill \\ \end{array}}}\right) \,\,\exp \frac{-i\,\,\left( {Et\,\,-\,\,\vec{p%
}\vec{x}}\right) }{\hbar }  \notag \\
&&  \notag \\
\psi _{D}^{(\ref{eq2})}\,\, &=&\,\,\left( {{\begin{array}{*{20}c} {U_2
\,\,\left( \vec {p} \right)} \hfill \\ {U_2 \,\,\left( \vec {p} \right)i}
\hfill \\ \end{array}}}\right) \,\,\exp \frac{-i\,\,\left( {Et\,\,-\,\,\vec{p%
}\vec{x}}\right) }{\hbar }  \notag \\
&&  \notag \\
\psi _{D}^{(\ref{eq3})}\,\, &=&\,\,\left( {{\begin{array}{*{20}c} {U_1
\,\,\left( \vec {p} \right)} \hfill \\ { - U_1 \,\,\left( \vec {p} \right)i}
\hfill \\ \end{array}}}\right) \,\,\exp \frac{-i\,\,\left( {Et\,\,-\,\,\vec{p%
}\vec{x}}\right) }{\hbar }  \label{eq26} \\
&&  \notag \\
\psi _{D}^{(\ref{eq4})}\,\, &=&\,\,\left( {{\begin{array}{*{20}c} {U_2
\,\,\left( \vec {p} \right)} \hfill \\ { - U_2 \,\,\left( \vec {p} \right)i}
\hfill \\ \end{array}}}\right) \,\,\exp \frac{-i\,\,\left( {Et\,\,-\,\,\vec{p%
}\vec{x}}\right) }{\hbar }  \notag
\end{eqnarray}

\noindent we have

\begin{equation}
\begin{array}{l}
\left( {\dfrac{E}{c}\,\,+\,\,\left\vert \vec{p}\right\vert }\right)
\,\,U_{1}\,\,-\,\,mc\,\,U_{2}\,\,=\,\,0 \\ 
\\ 
-\,\,mc\,\,U_{1}\,\,+\,\,\left( {\dfrac{E}{c}\,\,-\,\,\left\vert \vec{p}%
\right\vert }\right) \,\,U_{2}\,\,=\,\,0.%
\end{array}
\label{eq27}
\end{equation}%
The existence of non-trivial solutions is ensured by

\begin{equation*}
\frac{E^2}{c^2}\,\, - \,\,\left| \vec {p} \right|^2\,\, - \,\,m^2c^2\,\, =
\,\,0
\end{equation*}

\noindent and

\begin{equation}  \label{eq28}
U_1^{(1,2)} \,\, = \,\,\frac{mc^2}{E\,\, + \,\,c\left| \vec {p} \right|}%
U_2^{(1,2)} .
\end{equation}

Let $\,$%
\begin{equation*}
U_{2}^{(\ref{eq1})}\,=\left( {{\begin{array}{*{20}c} 1 \hfill \\ 0 \hfill \\
\end{array}}}\right) \text{ and}\,\ U_{2}^{(\ref{eq2})}\,=\left( {{%
\begin{array}{*{20}c} 0 \hfill \\ 1 \hfill \\ \end{array}}}\right) .
\end{equation*}

Then%
\begin{equation}
\begin{tabular}{l}
$\psi _{D}^{(\ref{eq1})}=\dfrac{mc^{2}N_{1}}{E+c\left\vert \vec{p}%
\right\vert }\left( {{\begin{array}{*{20}c} 1 \hfill \\ 0 \hfill \\
\end{array}}}\right) \otimes \left( {{\begin{array}{*{20}c} 1 \hfill \\ i
\hfill \\ \end{array}}}\right) \,\,\exp \dfrac{-i\,\,\left( {Et\,\,-\,\,\vec{%
p}\vec{x}}\right) }{\hbar }\,\,$ \\ 
$\ \ \ \ \ \ \ \ \ \ \ \ \ =\dfrac{mc^{2}N_{1}}{E+c\left\vert \vec{p}%
\right\vert }\left( {{\begin{array}{*{20}c} 1 \hfill \\ i \hfill \\ 0 \hfill
\\ 0 \hfill \\ \end{array}}}\right) \,\exp \dfrac{-i\,\,\left( {Et\,\,-\,\,%
\vec{p}\vec{x}}\right) }{\hbar }\,\,$ \\ 
\\ 
$\psi _{D}^{(\ref{eq2})}=\,\,N_{1}\left( {{\begin{array}{*{20}c} 0 \hfill \\
1 \hfill \\ \end{array}}}\right) \otimes \left( {{\begin{array}{*{20}c} 1
\hfill \\ i \hfill \\ \end{array}}}\right) \,\,\exp \dfrac{-i\,\,\left( {%
Et\,\,-\,\,\vec{p}\vec{x}}\right) }{\hbar }\,$ \\ 
$\ \ \ \ \ \ \ \ \ \ \ \ \ =N_{1}\left( {{\begin{array}{*{20}c} 0 \hfill \\
0 \hfill \\ 1 \hfill \\ i \hfill \\ \end{array}}}\right) \,\exp \dfrac{%
-i\,\,\left( {Et\,\,-\,\,\vec{p}\vec{x}}\right) }{\hbar }\,.$%
\end{tabular}
\label{eq29}
\end{equation}

Now

\begin{equation}
U_{2}^{(3,4)}\,\,=\,\,\frac{(E+c\left\vert \vec{p}\right\vert )}{mc^{2}}%
U_{1}^{(3,4)}.  \label{eq30}
\end{equation}

Let $\,$%
\begin{equation*}
U_{1}^{(\ref{eq3})}\,=\left( {{\begin{array}{*{20}c} 1 \hfill \\ 0 \hfill \\
\end{array}}}\right) \text{ and }\,U_{1}^{(\ref{eq4})}\,=\left( {{%
\begin{array}{*{20}c} 0 \hfill \\ 1 \hfill \\ \end{array}}}\right) .
\end{equation*}

Then

\begin{equation}
\begin{tabular}{l}
$\psi _{D}^{(\ref{eq3})}\,=N_{2}\left( {{\begin{array}{*{20}c} 1 \hfill \\ 0
\hfill \\ \end{array}}}\right) \otimes \left( {{\begin{array}{*{20}c} 1
\hfill \\ { - i} \hfill \\ \end{array}}}\right) \,\,\exp \dfrac{-i\,\,\left( 
{Et\,\,-\,\,\vec{p}\vec{x}}\right) }{\hbar }\,$ \\ 
$\ \ \ \ \ \ \ \ \ \ \ \ \ \ =N_{2}\left( {{\begin{array}{*{20}c} 1 \hfill
\\ { - i} \hfill \\ 0 \hfill \\ 0 \hfill \\ \end{array}}}\right) \,\exp 
\dfrac{-i\,\,\left( {Et\,\,-\,\,\vec{p}\vec{x}}\right) }{\hbar }\,$ \\ 
$\psi _{D}^{(\ref{eq4})}=\dfrac{(E+c\left\vert \vec{p}\right\vert )N_{2}}{%
mc^{2}}\left( {{\begin{array}{*{20}c} 0 \hfill \\ 1 \hfill \\ \end{array}}}%
\right) \otimes \left( {{\begin{array}{*{20}c} 1 \hfill \\ { - i} \hfill \\
\end{array}}}\right) \,\exp \dfrac{-i\,\,\left( {Et\,\,-\,\,\vec{p}\vec{x}}%
\right) }{\hbar }\,$ \\ 
$\ \ \ \ \ \ \ \ \ \ \ \ \ =\dfrac{(E+c\left\vert \vec{p}\right\vert )N_{2}}{%
mc^{2}}\left( {{\begin{array}{*{20}c} 0 \hfill \\ 0 \hfill \\ 1 \hfill \\ {
- i} \hfill \\ \end{array}}}\right) \exp \dfrac{-i\,\,\left( {Et\,\,-\,\,%
\vec{p}\vec{x}}\right) }{\hbar }$%
\end{tabular}
\label{eq31}
\end{equation}

Our choice is made in order to compare with the standard set of the linearly
independent solutions \noindent for the Dirac equation. Indeed, the
alternative%
\begin{equation}
\begin{tabular}{ll}
& $\psi _{D}^{(\ref{eq3})}=\dfrac{mc^{2}N_{1}}{E+c\left\vert \vec{p}%
\right\vert }\left( {{\begin{array}{*{20}c} 1 \hfill \\ { - i} \hfill \\ 0
\hfill \\ 0 \hfill \\ \end{array}}}\right) \,\exp \dfrac{-i\,\,\left( {%
Et\,\,-\,\,\vec{p}\vec{x}}\right) }{\hbar }$ \\ 
$\text{and}$ &  \\ 
& $\psi _{D}^{(\ref{eq4})}=N_{1}\left( {{\begin{array}{*{20}c} 0 \hfill \\ 0
\hfill \\ 1 \hfill \\ { - i} \hfill \\ \end{array}}}\right) \,\exp \dfrac{%
-i\,\,\left( {Et\,\,-\,\,\vec{p}\vec{x}}\right) }{\hbar }.$%
\end{tabular}%
\end{equation}

\noindent may serve us equally well and at the same time make things more
transparent, since we have \noindent obtained exactly the same solutions as $%
\psi _{D}^{(\ref{eq1})}$ and $\psi _{D}^{(\ref{eq2})}$, which would be
negative energy solutions in \noindent the Dirac equation with $\vec{p}%
^{\prime }$= -$\vec{p}$ if we make the substitution

\begin{equation}
i^{\prime }=-i=-\frac{e_{1}p_{1}+\,e_{2}p_{2}\,+\,e_{1}e_{2}p_{3}}{%
\left\vert \vec{p}\right\vert }.  \label{eq32}
\end{equation}

Obviously, the obtained set is mutually orthogonal.

Finally, using standard normalization condition, we obtain:%
\begin{eqnarray}
\psi _{D}^{(\ref{eq1})}\,\, &=&\frac{1}{2}\sqrt{\frac{mc^{2}}{E+c\left\vert 
\vec{p}\right\vert }}\left( {{\begin{array}{*{20}c} 1 \hfill \\ i \hfill \\
0 \hfill \\ 0 \hfill \\ \end{array}}}\right) \,\exp \frac{-i\,\,\left( {%
Et\,\,-\,\,\vec{p}\vec{x}}\right) }{\hbar }\,  \notag \\
&&\,  \notag \\
\psi _{D}^{(\ref{eq2})}\,\, &=&\frac{1}{2}\sqrt{\frac{E+c\left\vert \vec{p}%
\right\vert }{mc^{2}}}\left( {{\begin{array}{*{20}c} 0 \hfill \\ 0 \hfill \\
1 \hfill \\ i \hfill \\ \end{array}}}\right) \,\exp \frac{-i\,\,\left( {%
Et\,\,-\,\,\vec{p}\vec{x}}\right) }{\hbar }  \notag \\
&&  \notag \\
\psi _{D}^{(\ref{eq3})}\,\, &=&\frac{1}{2}\sqrt{\frac{mc^{2}}{E+c\left\vert 
\vec{p}\right\vert }}\left( {{\begin{array}{*{20}c} 1 \hfill \\ { - i}
\hfill \\ 0 \hfill \\ 0 \hfill \\ \end{array}}}\right) \exp \frac{%
-i\,\,\left( {Et\,\,-\,\,\vec{p}\vec{x}}\right) }{\hbar }  \label{eq33} \\
&&  \notag \\
\psi _{D}^{(\ref{eq4})}\,\, &=&\frac{1}{2}\sqrt{\frac{E+c\left\vert \vec{p}%
\right\vert }{mc^{2}}}\left( {{\begin{array}{*{20}c} 0 \hfill \\ 0 \hfill \\
1 \hfill \\ { - i} \hfill \\ \end{array}}}\right) \,\exp \frac{-i\,\,\left( {%
Et\,\,-\,\,\vec{p}\vec{x}}\right) }{\hbar }.  \notag
\end{eqnarray}

The obtained solutions maintain symmetry with respect to space coordinates
that may be expected \noindent based on the assumption of homogeneity of the
space-time continuum. Indeed, the correctness of the \noindent suggested
equations may be verified only through careful comparison with the
experimental data.

Now let us consider similar equations

\begin{equation}
\begin{array}{l}
\dfrac{1}{c}\,\,\dfrac{\partial \psi }{\partial t}\,\,i\,\,-\,\,\dfrac{%
\partial \psi }{\partial x_{j}}\,\,e_{j}\,\,=\,\,\dfrac{mc}{\hbar }\,\,\phi
\,i \\ 
\\ 
-\dfrac{1}{c}\,\,\dfrac{\partial \phi }{\partial t}\,\,i\,\,-\,\,\dfrac{%
\partial \phi }{\partial x_{j}}\,\,e_{j}\,\,=\,\,\dfrac{mc}{\hbar }\,\,\psi
\,i%
\end{array}%
\quad j=1,2,3  \label{eq34}
\end{equation}

\noindent and verify that they admit an additional set of plane wave
solutions, for example, in the following form:

\begin{equation}
\begin{array}{l}
\psi _{j}\,\,=\,\,U_{1}e_{j}\exp \dfrac{-e_{j}\,\,\left( {Et\,\,-\,\,\vec{p}%
\vec{x}}\right) }{\hbar }\, \\ 
\\ 
\phi _{j}\,\,=\,\,U_{2}\exp \dfrac{-e_{j}\,\,\left( {Et\,\,-\,\,\vec{p}\vec{x%
}}\right) }{\hbar }\,\,.%
\end{array}%
\quad j=1,2,3  \label{eq35}
\end{equation}

Here $U_1 $ and $U_2 $ are assumed to be real numbers. Then

\begin{equation}
\begin{array}{l}
U_{1}e_{j}\,\,\left( {\ -\dfrac{e_{j}Ei}{c}\,\,-\,\,e_{j}p_{1}e_{1}\,\,-\,%
\,e_{j}p_{2}e_{2}\,\,-e_{j}p_{3}e_{1}e_{2}}\right) \,\,=\,\,mcU_{2}i \\ 
\\ 
U_{2}\,\,\left( {\dfrac{e_{j}Ei}{c}\,\,-\,\,e_{j}p_{1}e_{1}\,\,-\,%
\,e_{j}p_{2}e_{2}\,\,-e_{j}p_{3}e_{1}e_{2}}\right) \,\,=\,\,mcU_{1}\,\,e_{j}i%
\end{array}
\label{eq36}
\end{equation}

\noindent or

\begin{equation}
\begin{array}{l}
U_{1}e_{j}\,\,\left( {-\dfrac{e_{j}Ei}{c}\,\,-\,\,e_{j}\,\,\left( {%
e_{1}p_{1}\,\,+\,\,e_{2}p_{2}\,\,+\,\,e_{1}e_{2}p_{3}}\right) }\right)
\,\,=\,\,mcU_{2}i \\ 
\\ 
U_{2}\,\,\left( {\dfrac{e_{j}Ei}{c}\,\,-\,\,e_{j}\,\,\left( {%
e_{1}p_{1}\,\,+\,\,e_{2}p_{2}\,\,+\,\,e_{1}e_{2}p_{3}}\right) }\right)
\,\,=\,\,mcU_{1}\,\,e_{j}i.%
\end{array}
\label{eq37}
\end{equation}

But according to (\ref{eq25})

\begin{equation}  \label{eq38}
e_1 p_1 \,\, + \,\,e_2 p_2 \,\, + \,\,e_1 e_2 p_3 \,\, = \,\,i\,\,\left| 
\vec {p} \right|.
\end{equation}

\newpage

\noindent which gives

\begin{equation}
\begin{array}{l}
U_{1}\,\,\left( {\dfrac{E}{c}\,\,+\,\,\left\vert \vec{p}\right\vert }\right)
i\,\,=\,\,mcU_{2}i \\ 
U_{2}e_{j}\,\,\left( {\dfrac{E}{c}\,\,-\,\,\left\vert \vec{p}\right\vert }%
\right) i\,\,=\,\,mcU_{1}e_{j}i%
\end{array}
\label{eq39}
\end{equation}

\noindent or

\begin{equation}
\begin{array}{l}
U_{1}\,\,\left( {\dfrac{E}{c}\,\,+\,\,\left\vert \vec{p}\right\vert }\right)
\,\,=\,\,mcU_{2} \\ 
U_{2}e_{j}\,\,\left( {\dfrac{E}{c}\,\,-\,\,\left\vert \vec{p}\right\vert }%
\right) \,\,=\,\,mcU_{1}e_{j}%
\end{array}
\label{eq40}
\end{equation}

\noindent and

\begin{equation}
\begin{array}{l}
U_{1}\,\,\left( {\dfrac{E}{c}\,\,+\,\,\left\vert \vec{p}\right\vert }\right)
\,\,=\,\,mcU_{2} \\ 
U_{2}\,\,\left( {\dfrac{E}{c}\,\,-\,\,\left\vert \vec{p}\right\vert }\right)
\,e_{j}\,=\,\,mcU_{1}e_{j}.%
\end{array}
\label{eq41}
\end{equation}

Thus, we finally obtain

\begin{equation}
\begin{array}{l}
\left( {\dfrac{E}{c}\,\,+\,\,\left\vert \vec{p}\right\vert }\right)
\,\,U_{1}\,\,-\,\,mcU_{2}\,\,=\,\,0 \\ 
-mc\,\,U_{1}\,\,+\,\,\left( {\dfrac{E}{c}\,\,-\,\,\left\vert \vec{p}%
\right\vert }\right) \,\,U_{2}\,\,=\,\,0%
\end{array}
\label{eq42}
\end{equation}

\noindent which justifies the above-made assumption concerning the reality
of $U_{1}$ and $U_{2}$. The existence of non-trivial solutions is ensured by

\begin{equation*}
\frac{E^2}{c^2}\,\, - \,\,\left| \vec {p} \right|^2\,\, - m^2c^2\,\, = \,\,0
\end{equation*}

\noindent and

\begin{equation}  \label{eq43}
U_1 \,\, = \,\,\frac{mc^2}{E\,\, + \,\,c\,\,\left| \vec {p} \right|}\,\,U_2
\end{equation}

Thus, we have obtained a triplet of solutions, each one associated with the
same mass, but with

\begin{equation}
\begin{array}{l}
\left[ {\psi _{j},\,\,\psi _{k}}\right] \,\,\neq \,0 \\ 
\left[ {\phi _{j},\,\,\phi _{k}}\right] \,\,\neq \,0\text{ \ \ \ \ \ \ }%
j,k\,\,=\,\,1,2,3 \\ 
\left[ {\psi _{j},\,\,\phi _{k}}\right] \,\,\neq 0\text{ \ \ \ \ \ \ }%
j\,\,\neq \,\,k%
\end{array}
\label{eq44}
\end{equation}

\newpage

In addition, in the capacity of $e_{j},\,\,j\,\,=\,\,1,2,3,$ one may choose
not only the quaternion basis itself but other sets, for example,

\begin{equation}
ie_{1}i,\text{ \ \ }ie_{2}i,\text{ \ \ }ie_{2}e_{1}i  \label{eq45}
\end{equation}

\noindent or

\begin{equation}
e_{1}ie_{1},\text{ \ \ }e_{2}ie_{2},\text{ \ \ }e_{1}e_{2}ie_{1}e_{2}.
\label{eq46}
\end{equation}

Notice, however, that (\ref{eq46}) do not form a quaternion but

\begin{equation}  \label{eq47}
e_1 ie_1 \,\, + e_2 ie_2 \,\, + \,\,e_1 e_2 ie_1 e_2 \,\, = \,\,i.
\end{equation}

Additional knowledge is required in order to define which set is relevant
and how it may be associated with the correspondent physical objects.

So far, our discussion has been restricted to the four-dimensional real
quadratic division algebra of quaternions. However, it is clear that the
equations (\ref{eq24}) and (\ref{eq34}) serve in an uniform manner also the
octonionic extension of the complex Hilbert space .

If the underlying algeraic foundation of the theory is extended to include

the eight-dimensional real quadratic division algebra of octonions, then the
corresponding additional set of solutions for the equations (\ref{eq34}) may
be obtained.

They may have, e.g., the following form:

\begin{eqnarray*}
\psi \,\, &=&\,\,U_{1}\,\,i\,\,\exp \frac{-i\,\,\left( {Et\,\,-\vec{p}\vec{x}%
}\right) }{\hbar }\, \\
\phi \,\, &=&\,\,U_{2}\,\,\exp \frac{-i\,\,\left( {Et\,\,-\vec{p}\vec{x}}%
\right) }{\hbar }\,
\end{eqnarray*}

($i$ is given by (\ref{eq25}))

\begin{equation}
\begin{array}{l}
{\psi _{k}\,\,=\,U_{1}e_{k}\exp }\dfrac{-e_{k}\,\,\left( {Et\,\,-\,\,\vec{p}%
\vec{x}}\right) }{\hbar }{\,} \\ 
{\phi _{k}\,\,=\,U_{2}\exp }\dfrac{-e_{k}\,\,\left( {Et\,\,-\,\,\vec{p}\vec{x%
}}\right) }{\hbar } \\ 
{\psi _{j_{k}}\,\,=U_{1}j_{k}\exp }\dfrac{-j_{k}\,\,\left( {Et\,\,-\,\,\vec{p%
}\vec{x}}\right) }{\hbar }{\,\,} \\ 
{\phi _{j_{k}}\,\,=\,U_{2}\exp }\dfrac{-j_{k}\,\,\left( {Et\,\,-\,\,\vec{p}%
\vec{x}}\right) }{\hbar }{\,\,}%
\end{array}%
\text{ \ \ \ \ \ \ }{k\,\,=\,\,4,5,6,7}  \label{eq48}
\end{equation}

\noindent where $U_1 ,\,\,U_2 $

\begin{equation*}
U_1 \,\, = \,\,\frac{mc^2}{E\,\, + \,\,c\left| \vec {p} \right|}\,\,U_2
\end{equation*}

\noindent are real numbers;

\begin{equation}
j_{k}\,\,=\,\,e_{k}i\,,\text{\ \ \ \ \ \ \ }k\,\,=\,\,4,5,6,7  \label{eq49}
\end{equation}

\noindent and, as before, form a quaternion. This quaternion turns out to
form an algebraic foundation of the momentum space and, therefore, the
algebraic symmetry between coordinate and momentum spaces may be broken in
this formulation.

The particularly symmetric case occurs, if $k\,\, = \,\,7$. Then

\begin{equation}
\begin{array}{l}
i\,\,=\,\,\dfrac{e_{1}p_{1}\,+\,\,\,e_{2}p_{2}\,\,+\,\,e_{3}p_{3}}{%
\left\vert \vec{p}\right\vert } \\ 
\\ 
j\,\,=\,\,\dfrac{e_{4}p_{1}\,+\,\,\,e_{5}p_{2}\,\,+\,\,e_{6}p_{3}}{%
\left\vert \vec{p}\right\vert }.%
\end{array}
\label{eq50}
\end{equation}

Indeed, in each case the set of equations (\ref{eq24}) and (\ref{eq34})
should be supplemented by corresponding leptonic equations, for example, for
(\ref{eq50})

\begin{equation}
\begin{array}{l}
\dfrac{1}{c}\,\,\dfrac{\partial \chi }{\partial t}\,\,j\,\,-\,\,\dfrac{%
\partial \chi }{\partial x}e_{4}\,\,-\,\,\dfrac{\partial \chi }{\partial y}%
e_{5}\,\,-\,\,\dfrac{\partial \chi }{\partial z}e_{6}\,\,=\,\,\dfrac{mc}{%
\hbar }\xi \\ 
\\ 
-\dfrac{1}{c}\,\,\dfrac{\partial \xi }{\partial t}\,\,j\,\,-\,\,\dfrac{%
\partial \xi }{\partial x}e_{4}\,\,-\,\,\dfrac{\partial \xi }{\partial y}%
e_{5}\,\,-\,\,\dfrac{\partial \xi }{\partial z}e_{6}\,\,=\,\,-\dfrac{mc}{%
\hbar }\chi .%
\end{array}
\label{eq51}
\end{equation}

Based on the results of M. Zorn \cite{17} that each automorphism of the
octonion algebra is completely defined by the images of three
\textquotedblleft independent\textquotedblright\ basis units \cite{18}, it
was demonstrated by M. G\"{u}naydin and F. G\"{u}rsey \cite{19} that under
given automorphism $\sigma $ we have three quaternionic planes in the space
formed by octonion algebra (space of quantum mechanical phases), which
undergo rotations by the angles $\phi _{1},\,\,\phi _{2},\,\,\phi _{3}$,
respectively, such that

\begin{equation}  \label{eq52}
\phi _1 \,\, + \,\,\phi _2 \,\, + \,\,\phi _3 \,\, = \,\,0\,\,\bmod \,\,2\pi
\end{equation}

\noindent remains invariant. The planes ($e_{i},\,\,e_{j})$ are determined
by the conditions $e_{i}e_{j}\,\,=\,\,e_{k}$ and $e_{k}$ is the fixed point
common to all of those planes (compare with (\ref{eq48}) and (\ref{eq49})).
These results might help to extract the set of independent solutions and to
obtain its correct classification.

It is worth mentioning that an alternative arrangement can be also possible.
We may consider a septet of solutions, each one associated with the same
mass, namely

\begin{equation}
\begin{array}{l}
\psi \,\,=\,\,U_{1}e_{k}\exp \dfrac{-e_{k}\,\,\left( {Et\,\,-\,\,\vec{p}\vec{%
x}}\right) }{\hbar }\,\, \\ 
\phi \,\,=\,\,U_{2}\exp \dfrac{-e_{k}\,\,\left( {Et\,\,-\,\,\vec{p}\vec{x}}%
\right) }{\hbar }\,\,%
\end{array}%
\text{ \ \ \ \ \ \ }k\,\,=\,\,1,...,7  \label{eq53}
\end{equation}

\noindent and, perhaps, an additional one in the form

\begin{equation}
\begin{array}{l}
\psi \,\,=\,\,U_{1}e_{j}\exp \dfrac{-e_{j}\,\,\left( {Et\,\,-\,\,\vec{p}\vec{%
x}}\right) }{\hbar }\, \\ 
\phi \,\,=\,\,U_{2}\exp \dfrac{-e_{j}\,\,\left( {Et\,\,-\,\,\vec{p}\vec{x}}%
\right) }{\hbar }\,%
\end{array}%
\text{ \ \ \ \ \ \ }j\,=\,\,1,...,7\,  \label{eq54}
\end{equation}

\noindent where

\begin{equation}
\begin{array}{l}
j_{k}\,\,=\,\,e_{k}ie_{k} \\ 
j_{k+3}\,\,=\,\,e_{k+3}\,\,je_{k+3} \\ 
j_{7}\,\,=\,\,e_{7}%
\end{array}%
\text{ \ \ \ \ \ \ }k\,\,=\,\,1,2,3  \label{eq55}
\end{equation}

\noindent or

\begin{equation}
\begin{array}{l}
j_{k}\,\,=\,\,ie_{k}i \\ 
j_{k+3}\,\,=\,\,je_{k+3}\,\,j \\ 
j_{7}\,\,=\,\,e_{7}%
\end{array}%
\text{ \ \ \ \ \ \ }k\,\,=\,\,1,2,3  \label{eq56}
\end{equation}

\noindent with the common fixed point $e_{7}$: $i$ and $j$ are given by (\ref%
{eq50}).

The number of independent solutions, which are arranged in such a way, is
sharply reduced and serves as a slight reminder of a similar possibility
discussed in the literature \cite{ 20}.

Now it may be clarified why we have chosen to discuss the Dirac equations in
the Weyl representation. The reason is merely technical. In order to perform
clean octonionic calculations, we assume that the solutions of the equations
(\ref{eq34}) have the form (\ref{eq35}), where $U_1 $ and $U_2 $ are real
numbers. Then the obtained relations (\ref{eq42}) justify the assumption.
Notice that the electroweak unification scheme \cite{ 15} is based on the
use of this representation of the Dirac equations; in addition, in that case
the solutions behave naturally with respect to the Lorentz transformations.

Let us demonstrate that the Dirac equations in the form (\ref{eq24}), as
well as the set (\ref{eq34}), permit a consistent probabilistic
interpretation (here the discussion is restricted to the case where the
underlying algebraic basis are quaternions).

\newpage

Consider formally

\begin{equation}
\begin{array}{l}
\dfrac{1}{c}\,\,\dfrac{\partial \psi }{\partial t}\,\,i\,\,-\,\dfrac{%
\partial \psi }{\partial x_{j}}e_{j}\,\,=\,\,\dfrac{m_{1}c}{\hbar }\phi
\,\,+\,\,\dfrac{m_{2}c}{\hbar }\phi i \\ 
\\ 
-\dfrac{1}{c}\,\,\frac{\partial \phi }{\partial t}\,\,i\,\,-\,\dfrac{%
\partial \phi }{\partial x_{j}}e_{j}\,\,=\,\,-\dfrac{m_{1}c}{\hbar }\psi
\,\,+\,\dfrac{m_{2}c}{\hbar }\psi i.%
\end{array}%
\quad j=1,2,3  \label{eq57}
\end{equation}

Then

\begin{equation}
\begin{array}{l}
\dfrac{1}{c}\,\,\dfrac{\partial \psi }{\partial t}\,\,\,\,+\,\,\dfrac{%
\partial \psi }{\partial x_{j}}e_{j}i\,\,\,=\,\,-\,\,\dfrac{m_{1}c}{\hbar }%
\phi \,i\,\,+\,\,\dfrac{m_{2}c}{\hbar }\phi \\ 
\\ 
\dfrac{1}{c}\,\,\dfrac{\partial \phi }{\partial t}\,\,\,-\,\,\dfrac{\partial
\phi }{\partial x_{j}}e_{j}i\,\,\,=\,\,-\dfrac{m_{1}c}{\hbar }\psi i\,\,-\,%
\dfrac{m_{2}c}{\hbar }\psi%
\end{array}%
\quad j=1,2,3  \label{eq58}
\end{equation}

\noindent and

\begin{equation}
\begin{array}{l}
\dfrac{1}{c}\,\,\dfrac{\partial \bar{\psi}}{\partial t}\,\,\,\,+\,\,ie_{j}%
\dfrac{\partial \bar{\psi}}{\partial x_{j}}\,\,\,=\,\,\,\dfrac{m_{1}c}{\hbar 
}i\bar{\phi}\,\,+\,\,\dfrac{m_{2}c}{\hbar }\bar{\phi} \\ 
\\ 
\dfrac{1}{c}\,\,\dfrac{\partial \bar{\phi}}{\partial t}\,\,\,\,-\,\,ie_{j}%
\dfrac{\partial \bar{\phi}}{\partial x_{j}}\,\,\,=\,\,\,\dfrac{m_{1}c}{\hbar 
}i\bar{\psi}\,\,-\,\,\dfrac{m_{2}c}{\hbar }\bar{\psi}.%
\end{array}%
\quad j=1,2,3  \label{eq59}
\end{equation}%
Thus, we have

\begin{equation}
\begin{array}{l}
\dfrac{1}{c}\,\,\dfrac{\partial \psi }{\partial t}\bar{\psi}\,\,\,\,+\,\,%
\dfrac{\partial \psi }{\partial x_{j}}e_{j}i\bar{\psi}\,\,\,=\,\,\,-\dfrac{%
m_{1}c}{\hbar }\phi \,i\bar{\psi}\,\,+\,\,\dfrac{m_{2}c}{\hbar }\phi \bar{%
\psi} \\ 
\\ 
\dfrac{1}{c}\,\,\psi \dfrac{\partial \bar{\psi}}{\partial t}\,\,\,+\,\,\psi
ie_{j}\,\,\dfrac{\partial \bar{\psi}}{\partial x_{j}}\,\,\,\,=\,\,\dfrac{%
m_{1}c}{\hbar }\psi i\bar{\phi}\,\,+\,\,\dfrac{m_{2}c}{\hbar }\psi \bar{\phi}
\\ 
\\ 
\dfrac{1}{c}\,\,\dfrac{\partial \phi }{\partial t}\bar{\phi}\,\,\,\,-\,\,%
\dfrac{\partial \phi }{\partial x_{j}}e_{j}i\bar{\phi}\,\,\,=\,\,\,-\dfrac{%
m_{1}c}{\hbar }\psi \,i\bar{\phi}\,\,-\,\,\dfrac{m_{2}c}{\hbar }\psi \bar{%
\phi} \\ 
\\ 
\dfrac{1}{c}\,\,\phi \dfrac{\partial \bar{\phi}}{\partial t}\,\,\,-\,\,\phi
ie_{j}\,\,\dfrac{\partial \bar{\phi}}{\partial x_{j}}\,\,\,=\,\,\dfrac{m_{1}c%
}{\hbar }\phi i\bar{\psi}\,\,-\,\,\dfrac{m_{2}c}{\hbar }\phi \bar{\psi}.%
\end{array}%
\quad j=1,2,3  \label{eq60}
\end{equation}

\newpage

Adding all equations (\ref{eq60}), we find

\begin{equation}
\begin{array}{l}
\dfrac{1}{c}\,\,\left\{ {\dfrac{\partial \psi }{\partial t}\bar{\psi}%
\,\,+\,\,\psi \dfrac{\partial \bar{\psi}}{\partial t}\,\,+\,\,\dfrac{%
\partial \phi }{\partial t}\bar{\phi}\,\,+\,\,\phi \dfrac{\partial \bar{\phi}%
}{\partial t}}\right\} \\ 
\\ 
+\,\,\left\{ {\dfrac{\partial \psi }{\partial x_{j}}e_{j}i\bar{\psi}%
\,\,+\,\,\psi ie_{j}\dfrac{\partial \bar{\psi}}{\partial x_{j}}\,\,-\,\,%
\dfrac{\partial \phi }{\partial x_{j}}e_{j}i\bar{\phi}\,\,-\,\,\phi
ie_{j}\,\,\dfrac{\partial \bar{\phi}}{\partial x_{j}}}\right\} =0%
\end{array}%
\quad j=1,2,3  \label{eq61}
\end{equation}

Notice that the mass terms vanish separately and, hence, the derivation
holds separately for (\ref{eq24}) and (\ref{eq34}) and from now on it is
understood that we consider the solutions of these equations separately.

The equation (\ref{eq61}) is invariant under the gauge transformation (\ref%
{eq15}) with $i$ given by (\ref{eq25}).

Then we have

\begin{equation}
\begin{array}{l}
\dfrac{1}{c}\,\,\dfrac{\partial }{\partial t}\,\,\,\,\psi \bar{\psi}\,\,+\,\,%
\dfrac{1}{c}\,\,\dfrac{\partial }{\partial t}\,\,\,\phi \bar{\phi} \\ 
\\ 
+\,\,\left\{ {\dfrac{\partial \psi }{\partial x_{j}}ie_{j}\bar{\psi}%
\,\,+\,\,\psi e_{j}i\,\,\dfrac{\partial \bar{\psi}}{\partial x_{j}}\,\,-\,\,%
\dfrac{\partial \phi }{\partial x_{j}}ie_{j}\bar{\phi}\,\,-\,\,\phi
e_{j}i\,\,\dfrac{\partial \bar{\phi}}{\partial x_{j}}}\right\} =0.%
\end{array}%
\quad j=1,2,3  \label{eq62}
\end{equation}

Adding equations (\ref{eq61}) and (\ref{eq62}), we obtain

\begin{equation}
\begin{array}{l}
\dfrac{1}{c}\,\,\dfrac{\partial }{\partial t}\,\,\left( {\psi \bar{\psi}%
\,\,\,\,+\,\,\,\phi \bar{\phi}}\right) \\ 
\\ 
+\,\,\dfrac{1}{2}{\LARGE \{}{\dfrac{\partial \psi }{\partial x_{j}}\left( {%
e_{j}i\,\,+\,\,ie_{j}}\right) \,\,\bar{\psi}\,\,\,\,+\,\,\psi \left( {%
e_{j}i\,\,+\,\,ie_{j}}\right) \,\,\dfrac{\partial \bar{\psi}}{\partial x_{j}}%
\,\,} \\ 
\\ 
{-\,\,\dfrac{\partial \phi }{\partial x_{j}}\,\,\left( {e_{j}i\,\,+\,\,ie_{j}%
}\right) \,\,\bar{\phi}\,\,-\,\,\phi \,\,\left( {e_{j}i\,\,+\,\,ie_{j}}%
\right) \,\,\dfrac{\partial \bar{\phi}}{\partial x_{j}}}{\LARGE \}}=0%
\end{array}%
\text{ \ \ \ \ \ \ }j=1,2,3  \label{eq63}
\end{equation}

\noindent or

\begin{equation}
\frac{\partial \rho }{\partial t}\,\,+\,\,\mathrm{div}\,\,\vec{j}\,\,=\,\,0
\label{eq64}
\end{equation}

\noindent where

\begin{equation}
\begin{array}{l}
\rho \,\,=\,\,\psi \bar{\psi}\,\,+\,\,\phi \bar{\phi} \\ 
\\ 
j_{k}\,\,=\,\,\dfrac{c}{2}\left\{ {\,\psi \left( {e_{k}i\,\,+\,\,ie_{k}}%
\right) \,\,\bar{\psi}\,\,-\,\,\phi \,\,\left( {e_{k}i\,\,+\,\,ie_{k}}%
\right) \,\,\bar{\phi}}\right\} .%
\end{array}%
\quad \text{\ \ \ }k=1,2,3  \label{eq65}
\end{equation}

If $i$ is given by (\ref{eq25}), then

\begin{equation}
e_{k}i\,\,+\,\,ie_{k}\,\,=\,\,-\frac{2p_{k}}{\left\vert \vec{p}\right\vert }%
,\quad k=1,2,3.  \label{eq66}
\end{equation}

Thus, in accordance with the Galileo, Maxwell and Schr\"{o}dinger theories,
the probability current $\vec{j}$ is proportional to the velocity operator,
which is a constant of motion for free particles.

\textit{Zitterbewegung }[3] phenomenon is absent in this formulation.

\section{Conclusion}

In account of the experimental information that became available during the

last century, it is desirable for the equations of motion for the
fundamental fermions to have the following properties:\smallskip

\noindent

- the equations should possess $SU\left( 2\right) \,\,\otimes U\left(
1\right) $ local gauge invariance \linebreak \hspace*{0.340in}intrinsically
since the electron is not a source of pure electromagnetic \linebreak 
\hspace*{0.340in}radiation but also has the ability to participate in weak
interactions;

\noindent

- the electron is the only particular member of the entire family of fun-

\ \ damental fermions, it is desirable that all fermions are described in the 

\ \ uniform manner;

\noindent

- there exist three replication of the families of the fundamental fermions;

\noindent

- leptons do not have a color;

\noindent

- quarks do have a color;

\noindent

- each quark appear in triplet associated with the same mass;

\noindent

- a color is associated with the internal degree of freedom;

\noindent

- a color symmetry can't be broken.

\smallskip

In contrast to the existing quaternionic formulations [9, 10] of the Dirac
equation, we suggested here closely related but essentially different sets
of equations that allow description of the free motion for electron and
neutrinos, as well as triplets of quarks. The suggested solution possesses $%
SU\left( 2\right) \,\,\otimes U\left( 1\right) $ local gauge invariance
intrinsically. All obtained solutions have the structure (A1) and (A2).
These solutions are substantially different from the standard ones and may
yield, therefore, different values for observable physical quantities. We
propose to compare and verify them against the existing experimental
information. The suggested reexamination may help to decide what is a
relevant mathematical framework suitable to achieve a solution of the
unification problem for the fundamental interactions.

During the last decades, an enormous progress in the understanding of
quantum theory of fields took place. It became almost apparent that we have
to deal with four essentially fundamental interactions, which have a similar
origin, namely, presence of phases in the quantum mechanical description of
the fundamental sources of these fields [21]. In addition, the investigation
of the properties of the fundamental sources (leptons, quarks) clearly
established that the quantum numbers in addition to electric charge (week
hypercharge and color) appear in amazing correspondence with the complex
variety of the radiated fields.

The formulation of classical mechanics and the classical theory of fields
have demonstrated that the presence of additional interactions requires a
suitable generalization of the mathematical language used. Application of
the analogy with the structure of classical physics in the framework of
functional analysis naturally concentrated around attempts to extent it on
all Hurwitz algebras, as the underlying algebraic foundation of the theory.

The above discussion may be considered as an additional step towards
realization of a program initiated by E. Schr\"{o}dinger [22] to treat all
of the physics as wave mechanics:

\smallskip

a) The universal mathematical architecture of the physics is given in terms
of ten functional - analytical frameworks, suitable to incorporate the
results of physical measurements.

Real, complex, quaternion and octonion states with real scalar product
should be equivalent to the theory of classical fields. Unification of
electromagnetism with gravitation should occur already in the classical
field theory.

Complex, quaternion and octonion states with complex scalar product should
allow realization of present unification schemes. Notice that pure
relativistic quantum electrodynamics does not exist because there are no
elementary sources of pure electromagnetic radiation. Neutrino is an
elementary source of pure weak radiation.

Quaternion and octonion states with quaternion scalar product should
describe wave mechanics of space-time continuum.

Octonion states with octonion scalar product should allow ultimate
realization of idea of elementary particles picture of natural phenomena.

\smallskip

b) One expects that the quantum mechanical space-time continuum should be
different from its classical counterpart. Perhaps, the spin is not a
dynamical variable, but the feature of the quantum mechanical world, namely,
the world point is described by the following expression (before inclusion
of quantum gravity):

\begin{equation*}
X\,\, = \,\,\left( {{\begin{array}{*{20}c} t \hfill & { - e_1 x\,\, -
\,\,e_2 y\,\, - \,\,e_3 z} \hfill \\ {e_1 x\,\, + \,\,e_2 y\,\, + \,\,e_3 z}
\hfill & t \hfill \\ \end{array} }} \right)
\end{equation*}

It is interesting to construct the correspondent metric space and the
application of Least Action should lead to the equations of motion for the
fundamental fermions.

\smallskip

c) In the present discussion we consider masses as external phenomenological
parameters. However, the structure of octonion quantum mechanics with
complex scalar product suggest natural mechanism to generate masses of the
fundamental fermions as energy gaps obtained after splitting states of
initially degenerated two-level physical system.

\bigskip

I am grateful to L.P. Horwitz, Y. Aharonov, S. Nussinov, and I.D. Vagner for
the stimulating discussions.

This work has been supported in part by the Binational Science Foundation,
Jerusalem.

\section*{Appendix}

This paper is concerned with the relativistic dynamics of single particle
states and for this reason we have used only the following results of the
particular realization of this program for the quaternionic and octonionic
Hilbert spaces with complex scalar products:

1) In the quaternionic extension, quantum mechanical states are represented
by

\begin{equation}
\begin{tabular}{ll}
& $\Psi \,\,_{\not\subset }^{\left( 1\right) \,}\,=\,\dfrac{1}{\sqrt{2}}%
\binom{f}{fe_{1}}$ \\ 
or &  \\ 
& $\Psi \,\,_{\not\subset }^{\left( 2\right) \,}\,=\,\dfrac{1}{\sqrt{2}}%
\binom{f}{-fe_{1}}$%
\end{tabular}
\tag{A1}  \label{A1}
\end{equation}

\noindent where $f\,\,=\,\,f_{0}\,\,+\,\,\sum\limits_{i\,\,=\,\,1}^{3}{%
f_{i}e_{i};\,\,\,\,f_{0},\,\,f_{i}\,\,}$ are real functions of the
space-time coordinates and $e_{i},\,\,i\,\,=\,\,1,\,\,2,\,\,3$ form a basis
for the real quadratic division algebra of quaternions;

In the octonionic extension quantum mechanical states are represented~by:

\begin{equation}
\begin{tabular}{ll}
& $\Psi \,\,_{\not\subset }^{\left( 3\right) \,}\,=\,\dfrac{1}{\sqrt{2}}%
\binom{f}{fe_{7}}$ \\ 
or &  \\ 
& $\Psi \,\,_{\not\subset }^{\left( 4\right) \,}\,=\,\dfrac{1}{\sqrt{2}}%
\binom{f}{-fe_{7}}$%
\end{tabular}
\tag{A2}  \label{A2}
\end{equation}

\noindent where $f\,\,=\,\,f_{0}\,\,+\,\,\sum\limits_{i\,\,=\,\,1}^{7}{%
f_{i}e_{i};\,\,\,\,f_{0},\,\,f_{i}\,\,}$are real functions of the space-time
coordinates and $e_{i}\,\,,\,\,i\,\,=\,\,1,\,\,...,\,\,7$ are a basis for
the real quadratic division algebra of octonions.

The $e_{1}$ and $e_{7}$ in the definition of the states (A1) and (A2) play
the role of a label for the generator of a complex field in the space of
one-body states. For example, any one of the quaternionic units or some
linear combination of them%
\begin{equation}
i\,\,=\,\,\frac{ae_{1}\,\,+\,\,be_{2}\,\,+\,\,ce_{3}}{\sqrt{%
a^{2}\,\,+\,\,b^{2}\,\,+\,\,c^{2}}}  \tag{A3}  \label{A3}
\end{equation}%
($a,b,c$ are arbitrary real numbers) may be used for this purpose. Thus, a
definition of this combination cannot be obtained kinematically and turns
out to be a matter of the dynamics of single particle states.

2) Consider the general form of operators, induced by the structure (A1) of
the vector space.

For the complex linear operators%
\begin{equation}
A_{z}\,\,=\,\,\left( {{\begin{array}{*{20}c} {a_{11} } \hfill & {a_{12} }
\hfill \\ {a_{21} } \hfill & {a_{22} } \hfill \\ \end{array}}}\right) 
\tag{A4}  \label{A4}
\end{equation}

\noindent where matrix elements $a_{ij}$ are real operators over quaternions
and, in turn, are assumed to be at least z-linear operators, we have

\begin{equation}
\Psi \,\,_{\not\subset }^{\left( 1\right) \,}\,^{\prime }=\,\frac{1}{\sqrt{2}%
}\,\left( {{\begin{array}{*{20}c} {f'} \hfill \\ {f'e_1 } \hfill \\
\end{array}}}\right) \,=\,\,A_{z}\Psi _{\not\subset }^{\left( 1\right) }=\,\,%
\frac{1}{\sqrt{2}\,}\left( {{\begin{array}{*{20}c} {a_{11} } \hfill &
{a_{12} } \hfill \\ {a_{21} } \hfill & {a_{22} } \hfill \\ \end{array}}}%
\right) \left( {{\begin{array}{*{20}c} f \hfill \\ {fe_1 } \hfill \\
\end{array}}}\right)  \tag{A5}  \label{A5}
\end{equation}

\noindent and%
\begin{equation}
a_{21}f\,+\,a_{22}f_{1}e_{1}\,=\,\,f^{\prime }e_{1}=\,\,\left( {%
a_{11}f\,\,+\,\,a_{12}fe_{1}}\right) \,\,e_{1}=\,a_{11}fe_{1}\,-\,\,a_{12}f.
\tag{A6}  \label{A6}
\end{equation}

Therefore,%
\begin{equation}
a_{12}\,\,=\,\,-a_{21};\,\,\,\,\,a_{11}\,\,=\,\,a_{22}.  \tag{A7}  \label{A7}
\end{equation}

The restrictions (A7) on the matrix elements of the operator (A4), obtained
for the states of the form $\Psi _{\not\subset }^{(\ref{eq1})}$ , are also
valid if one considers the transformations%
\begin{equation}
\Psi _{\not\subset }^{(\ref{eq2})\,^{\prime }}=\,A_{z}\,\,\Psi _{\not\subset
}^{(\ref{eq2})}.  \tag{A8}  \label{A8}
\end{equation}

However, for the operators transforming the state $\Psi _{\not\subset }^{(%
\ref{eq1})}$ into the state $\Psi _{\not\subset }^{(\ref{eq2})}$ (and vice
versa),%
\begin{equation}
\Psi _{\not\subset }^{\left( 2\right) \,\,\,\,^{\prime }}\,\,=\,\,\frac{1}{%
\sqrt{2}}\,\,\left( {{\begin{array}{*{20}c} {f'} \hfill \\ { - f'e_1 }
\hfill \\ \end{array}}}\right) \,\,=\,\,\,\,\left( {{\begin{array}{*{20}c}
{a_{11} } \hfill & {a_{12} } \hfill \\ {a_{21} } \hfill & {a_{22} } \hfill
\\ \end{array}}}\right) \,\,\,\frac{1}{\sqrt{2}}\,\,\,\left( {{%
\begin{array}{*{20}c} f \hfill \\ {fe_1 } \hfill \\ \end{array}}}\right) , 
\tag{A9}  \label{A9}
\end{equation}

\noindent we have%
\begin{equation}
a_{21}\,\,=\,\,a_{12};\,\,\,\,\,a_{11}\,\,=\,\,-a_{22}.  \tag{A10}
\label{A10}
\end{equation}

Thus, we have obtained two possible types of complex linear operators,
either 
\begin{equation}
{\begin{array}{*{20}c} \hfill & {A_z^{(\ref{eq1})} \,\, = \,\,\left(
{{\begin{array}{*{20}c} {a_{11} } \hfill & {a_{12} } \hfill \\ { - a_{12} }
\hfill & {a_{11} } \hfill \\ \end{array} }} \right)} \hfill & \hfill
\end{array}}  \tag{A11}  \label{A11}
\end{equation}%
or 
\begin{equation}
{\begin{array}{*{20}c} \hfill & {A_z^{(\ref{eq2})} \,\, = \,\,\left(
{{\begin{array}{*{20}c} {a_{11} } \hfill & {a_{12} } \hfill \\ {a_{12} }
\hfill & { - a_{11} } \hfill \\ \end{array} }} \right)} \hfill & \hfill
\end{array}}  \tag{A12}  \label{A12}
\end{equation}

We remark that the matrix elements $a_{11} $ and $a_{12} $ do not commute.


\begin{thebibliography}{99}
\bibitem{18} By three "independent" elements is implied any three elements $%
e_{i},\,\,e_{j}\,\,$and$\,\,e_{k}$ such that none of them is proportional to
a product of the other two, i.e. $e_{k}\,\,\neq \,\,ae_{i}e_{j}.$
\end{thebibliography}
\end{document}